\documentclass[aps,prl,twocolumn,groupedaddress]{revtex4}

\usepackage{graphicx}
\usepackage{amsmath}
\usepackage{bbold}
\begin{document}
\title{Surface state charge dynamics of a high-mobility three dimensional topological insulator}


\author{Jason N. \surname{Hancock}$^{1}$, J. L. M. van Mechelen$^{1}$, Alexey B. Kuzmenko$^{1}$, Dirk van der Marel$^{1}$, Christoph Br\"une$^{2}$, Elena G. Novik$^{2}$, Georgy V. Astakhov$^{2}$, Hartmut Buhmann$^{2}$, Laurens W. Molenkamp$^{2}$}
\affiliation{$^{1}$
D\'epartement de Physique de la Mati\`ere Condens\'ee, Universit\'e de Gen\`eve, quai Ernest-Ansermet 24, CH 1211 Gen\`eve 4, Switzerland}
\affiliation{$^{2}$
Physikalisches Institut der Universit\"at W\"urzburg - 97074 W\"urzburg, Germany}

\date{\today}
\begin{abstract}
We present a magneto-optical study of the three-dimensional topological insulator, strained HgTe using a technique which capitalizes on advantages of time-domain spectroscopy to amplify the signal from the surface states. This measurement delivers valuable and precise information regarding the surface state dispersion within $<$1 meV of the Fermi level. The technique is highly suitable for the pursuit of the topological magnetoelectric effect and axion electrodynamics.
\end{abstract}
\maketitle
Spin 1/2 particles exhibit a counterintuitive property that their wavefunction acquires a $\pi$ phase upon 360$^\circ$ rotation. If spin and orbital degrees of freedom are mixed in a particular way, the momenta of electrons in a crystalline lattice feel important effects of this $\pi$ `Berry's phase', which can lead to a new phase of matter whose description requires a fundamental redress of the theory of semiconductors \cite{kanemele05,zhang09,hasankanereview10}, and probably many other materials classes \cite{raghu08,groth09,dzero10}. These topological insulators exhibit an odd number of metallic surface bands with helical spin texture \cite{hsieh08,hsieh09,chen09} surrounding the nominally insulating bulk and display characteristic suppression of backscattering from step edges and nonmagnetic impurities \cite{alpinchshev10,roushan09}. We present magneto-optical measurements deep in the terahertz frequency regime exploring the charge dynamics of surface states in high-mobility strained films of HgTe \cite{fu07,dai08,luo10,brune11}. Using a time-domain technique, we detect a strong magneto-optical signal which is dominated by surface bands. This information reveals precise details of the low-energy excitations and momentum-energy dispersion of the helical metallic surface state.

The observation of the quantum spin Hall effect (QSHE) in CdTe-HgTe-CdTe quantum wells \cite{bernevig06,konig07} represents a significant advance in the ability to robustly segregate electronic currents of opposite spin, an effect which paves the way to new applications for spintronics and fault-tolerant quantum computation. In bulk, HgTe possesses the band inversion property due to spin-orbit interaction, a prerequisite condition for the QSHE, but also needed to realize a three dimensional topological insulator phase. Unfortunately, the Fermi level appears directly at the intersection of two bands, rendering the bulk semimetallic and leaving the surface states and the expected topological aspects of this material obscured by the low-energy bulk excitations. However, when HgTe is compressively strained against a CdTe substrate, the bulk band intersection becomes fully gapped in response to the lowered symmetry, permitting isolated access to helical surface bands \cite{fu07,dai08,luo10,brune11}. Recently, strong evidence for the existence of the $\pi$ Berry's phase, surface bands, and associated zero Landau level were observed as a quantum Hall effect of surface states, whose existence were subsequently verified using angle-resolved photoemission \cite{brune11}.

\begin{figure}
\begin{center}
\includegraphics[width=3.5in]{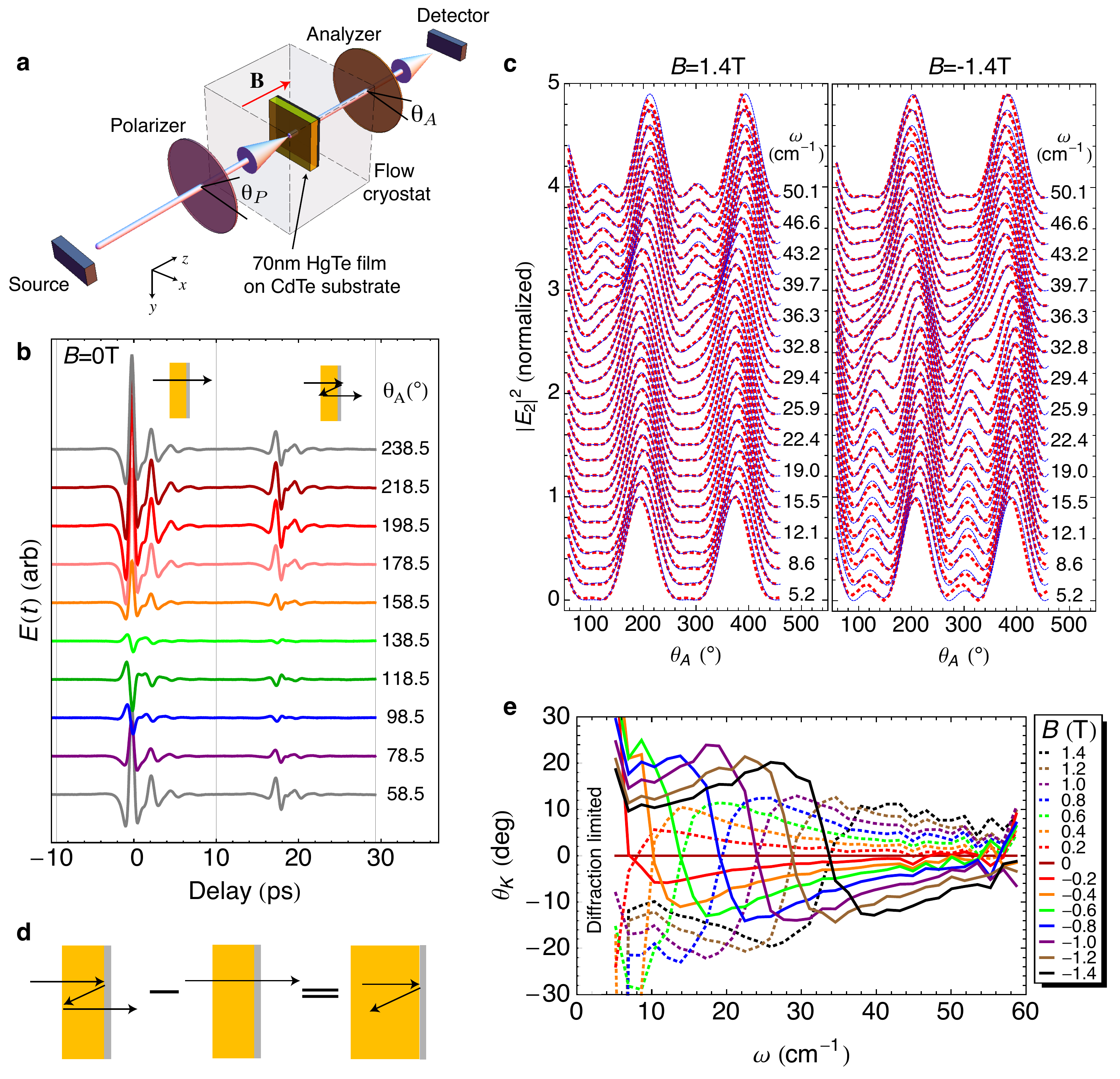}
\caption{(Color online) Time-domain measurement of the normal-incidence Kerr angle. (a) Illustration of the magneto-optical THz apparatus. $\theta_P$ was set to 51.2$^\circ$ throughout. (b) Subset of the angle-dependent measurements of the 1st and 2nd pulses. Vertical lines show the cuts used to separate the pulses into $E_1$ and $E_2$. (c) Measured $|E_2|^2$ (red dashes) and the fit from eqn (\ref{eq:Ei}) (blue lines) used to determine the Faraday angle $\theta_{F,2}(\omega)$ for +1.4T and -1.4T. (d) Pictorial rationale for the expression $\theta_{F,2}-\theta_{F,1}=\theta_K$. (e) The measured $\theta_K$ for several magnetic field values.}
\label{fig:MO}
\end{center}
\end{figure}

In order to probe directly the surface states of strained HgTe, we have developed a method of acquiring time-domain terahertz magneto-optical data using a home-built superconducting magnet in a flow cryostat, illustrated in Figure \ref{fig:MO}a. Complementary to frequency-domain measurements performed on the same sample at higher temperature ($>$50 K) \cite{shuvaev11}, which were interpreted in terms of a thermally-activated carriers in the bulk, our measurements were performed at low temperature, where the terahertz response is due to a combination of intrinsically doped bulk carriers and topological surface bands \cite{brune11}. At 4.35 K, the magnetic field and incident polarizer angle $\theta_P$ were set, and one direct pulse plus one echo, due to internal reflection inside the substrate, were collected for analyzer angles $\theta_A$ subtending 400$^\circ$ in 10$^\circ$ steps. The pulse and echo were then partitioned into $E_1(t,\theta_A,B)$ and $E_2(t,\theta_A,B)$ as shown in Figure 1b, and separately Fourier transformed to obtain the electric field amplitudes $E_1(\omega,\theta_A,B)$ and $E_2(\omega,\theta_A,B)$.

When the pulse passes through or reflects from the film, the electric field direction can rotate due to off-diagonal elements of the dielectric tensor or possibly the predicted topological magneto-electric effect \cite{qhz08}. Accounting for the horizontally-polarizing detector and emitter antenna of the spectrometer, the electric field amplitude in pulse $i$ at the detector is
\begin{equation}
\label{eq:Ei}
E_{i}=E_s\cos\theta_P\cos\theta_A\cos(\theta_A-\theta_P+\theta_{F,i})
\end{equation}
where $E_s$ is the electric field generated by the source and $\theta_{F,i}$ is the complex, sample-induced Faraday rotation of the $i$th pulse. Figure \ref{fig:MO}d,e shows an example fit of equation (\ref{eq:Ei}) to the measured intensity $|E_2(\omega,\theta_A,B)|^2$ of the second pulse for fields of $\pm$1.4 T. The change in shape of the intensity distribution around 35 $cm^{-1}$ is caused by a change in the complex Faraday angle $\theta_{F,2}$ at this frequency, which we attribute to an absorptive transition among surface state Landau levels. The fitting procedure was applied to both pulses to uniquely determine $\theta_{F,1}$ and $\theta_{F,2}$ at each frequency and field value measured. 

\begin{figure}
\begin{center}
\includegraphics[width=3.5in]{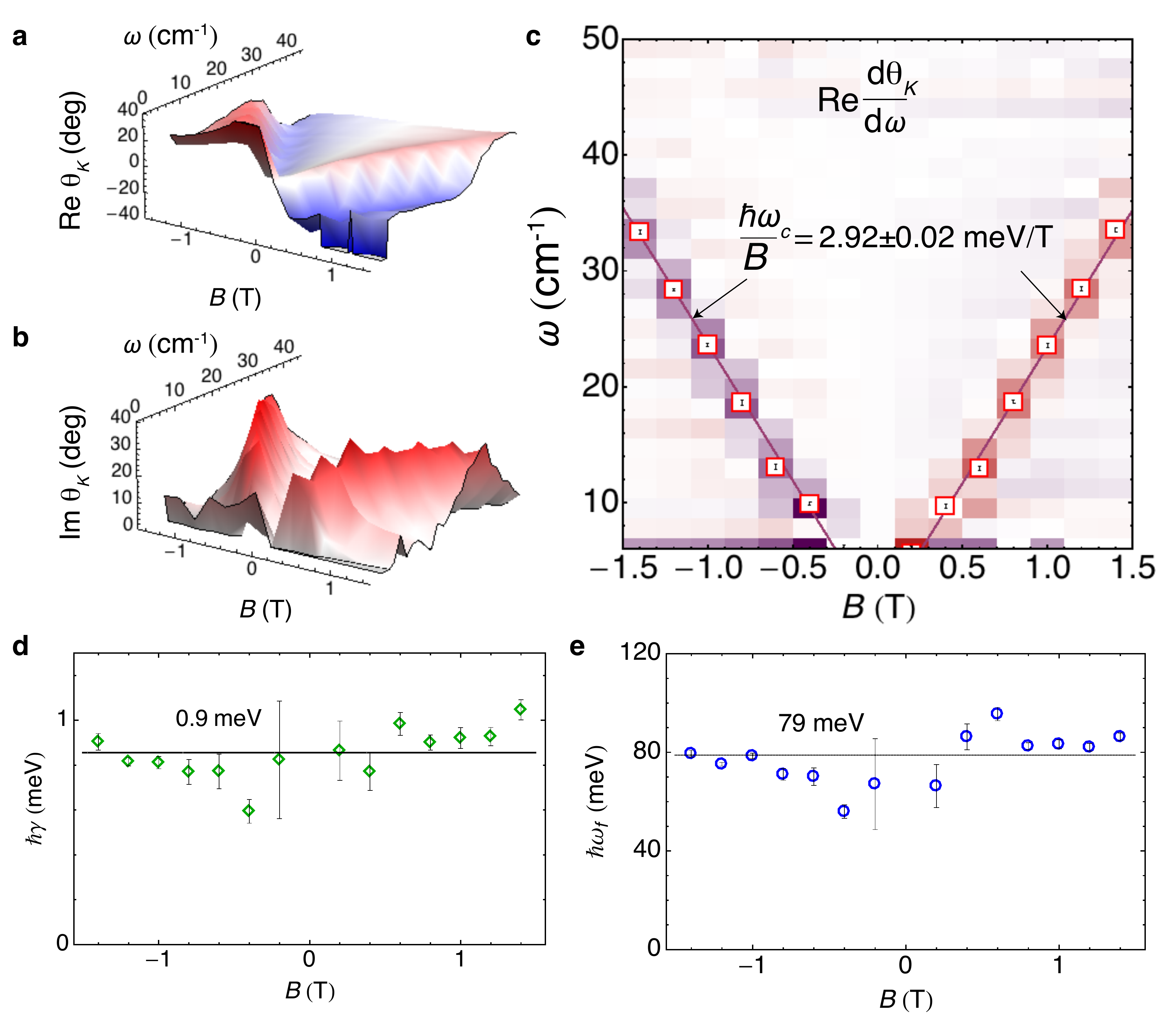}
\caption{(Color online) Properties of the surface-state cyclotron resonance. Real (a) and imaginary (b) Kerr angle $\theta_K(\omega,B)$ determined from terahertz ellipsometry. (c) The experimentally determined $\omega_c$ (red squares) and a linear fit (purple lines) showing the $B$-linear CR behavior. $\omega_c$ corresponds to the inflections in Re$\theta_K(\omega)$, as demonstrated by the experimental frequency derivative Re$d\theta_K/d\omega$, shown as a false color plot in the background. (d,e) Fit parameters $\gamma$ and $\omega_f$. The error bars arise from scatter in the determined CR parameters.}
\label{fig:CR}
\end{center}
\end{figure}

The simultaneous acquisition of multiple pulses with separate histories of contact with the film provides an advantage in accurately determining the polarization rotation due to the film. Any polarizing effects of the polyethylene cryostat windows, as well as inaccuracy in repositioning the polarizer following an angle sweep is exactly the same for the two pulses, so taking the difference in the two Faraday angles $\theta_K$=$\theta_{F,2}-\theta_{F,1}$ cancels these extraneous effects\footnote{This interpretation of the difference angle is possible because the CdTe substrate showed negligible rotation under similar conditions.}. A separate measurement of a reference CdTe substrate showed negligible rotation in magnetic field, and this normal-incidence Kerr angle $\theta_K$, shown in Figure \ref{fig:MO}e and \ref{fig:CR}a-b is precisely the rotation of polarization induced when the pulse reflects at normal incidence from the substrate side of the film (Fig. \ref{fig:MO}d), and the accuracy to which it can be determined is greater than for the Faraday angles separately.

The field-induced changes in $\theta_K$ are typical of cyclotron resonance (CR) behavior, for which the dynamical conductivity appropriate for photon angular momentum either parallel (+) or antiparallel (-) the momentum of the incident photon is:
\begin{equation}
\label{eq:sigma}
\sigma_{\pm}(\omega)=i\frac{e^2}{h}\sum_{j}\frac{\omega^j_{f}}{\omega\pm\omega^j_c+i\gamma}
\end{equation}
where $\omega_c^j$ is the cyclotron frequency, $\omega_f^j$ the Drude weight, and $\gamma^j$ is the inverse lifetime of the $j$th band. The elementary response functions in Equation (\ref{eq:sigma}) are then plugged into the expression
\begin{equation*}
\label{ }
\tan{\theta_K}=-\frac{iZ_0n_s(\sigma_+-\sigma_-)}{n_s^2-(1+Z_0\sigma_+)(1+Z_0\sigma_-)}
\end{equation*}
to fit the measured Kerr angle. Here, $n_s$ is the measured substrate index of refraction and $Z_0$ is the impedance of free space (see Supplemental Materials \footnote{See Supplemental Material for the
measured refractive index $n_s(\omega)$ of CdTe, quality of fits to
$\theta_K$, and derivation of relevant equations presented in the text.}). While we explore the possibility of a multi-component CR below, the resonance at each field is fit and satisfactorily described by a single effective $\omega_c$, $\gamma$, and $\omega_f$. 
This procedure results in the parameters summarized in Figures \ref{fig:CR}c-e. The sign of the rotation indicates that the carriers involved in the CR are electron-like, and fits to the field dependence give $\hbar\gamma=0.9$ meV, a field-linear cyclotron frequency with $\hbar\omega_c/B=2.92\pm 0.02$ meV/T, and total Drude weight $\hbar\omega_f=79.0\pm 1.7$ meV. Using the relation $\mu B = \omega_c/\gamma$, and the nearly field-independent $\gamma$ (Figure \ref{fig:CR}d), these fits imply a very high carrier mobility $\mu=34,220$ cm$^{2}$/V$\cdot$ s, consistent with previous transport \cite{brune11} and optical \cite{shuvaev11} measurements. 

In the limit of small carrier concentration, which as we will see below is the relevant limit in the context of this discussion, the 2D Fermi-surfaces are isotropic, and $\omega_f^j$ and $\omega_c^j$ at low fields are uniquely determined by the Fermi momentum, $k_{F}^j$, and the group velocity at the Fermi-energy, $v_{F}^j$ (see Supplemental Materials)
\begin{eqnarray*}
\label{ }
\omega_{f}^j&=&\frac{1}{2}v_{F}^jk_{F}^j \\ \nonumber
\omega_{c}^j&=&\frac{eB}{\hbar}\frac{v_{F}^j}{k_{F}^j}.
\end{eqnarray*}
Knowledge of $\omega_f^j$ and $\omega_c^j$ therefore permits determination of the parameters $v_{F}^j$ and $k_{F}^j$, which in turn provides valuable and precise information on the band structure at and around the Fermi energy.

In our 70 nm films, the conduction band is quantized and well-separated in energy due to confinement in the $z$ direction, and at low temperature any possible bulk contributions come from a small number of two-dimensional electron pockets. Unlike bulk, the surface contributions to the Drude weight are present for all values of the chemical potential, due to their gapless nature. Because the surface bands are at higher filling ($k_{F}^s>k_F^b$) and are more steeply dispersing ($v_{F}^s>v_F^b$) than the conduction bands, these states not only always contribute to the CR, but also always contribute more strongly to the magneto-optical signal than a set of bulk states with the same Fermi surface area. 

\begin{figure}
\begin{center}
\includegraphics[width=3.5in]{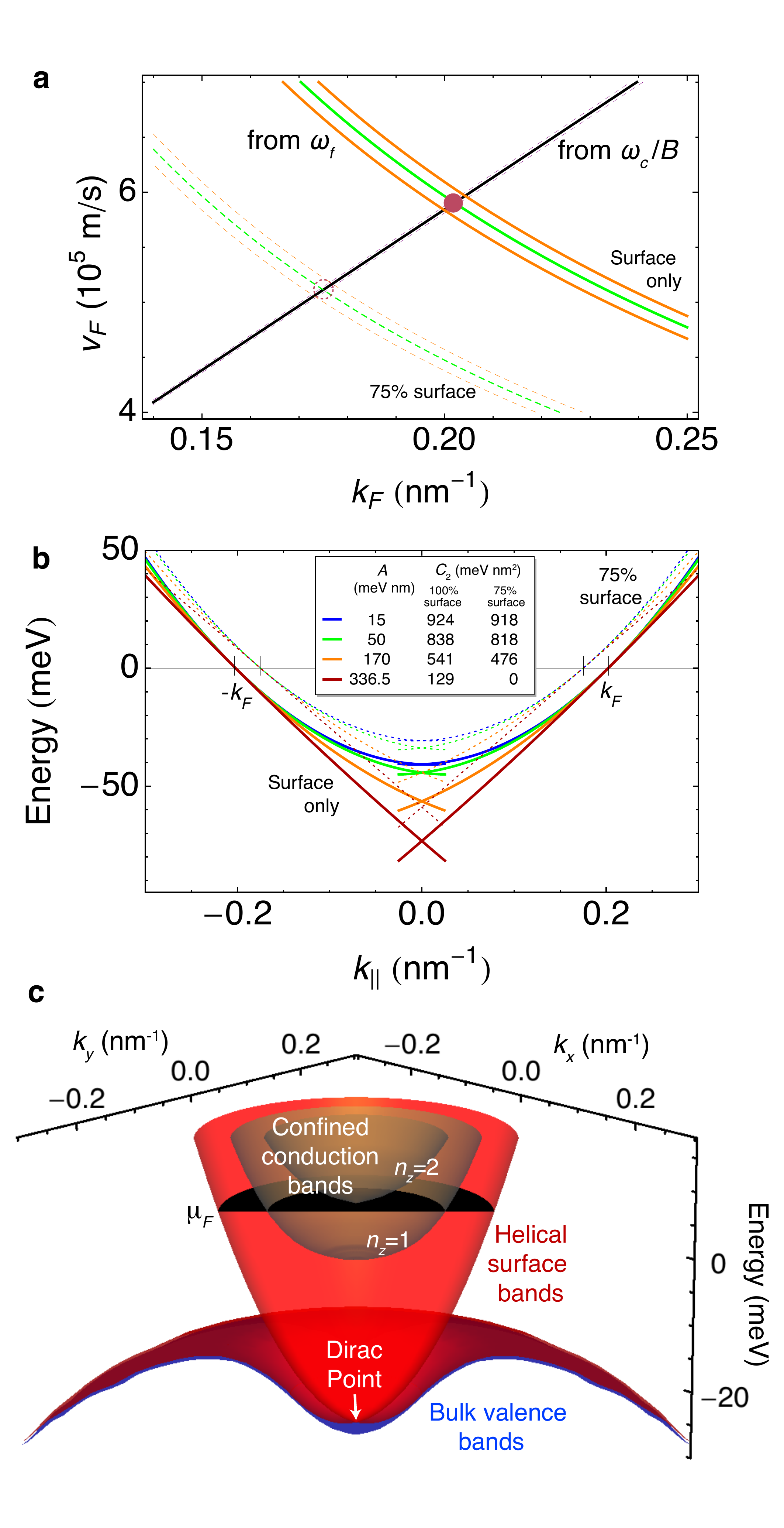}
\caption{(Color online) Dispersion of the surface state band. (a) Constraints on Fermi parameters $k_F$ and $v_F$ consistent with the magneto-optical data. Black line shows the constraint from the observed $\omega_c/B$, and purple dashes indicate the error bars on this quantity. Green solid curve is a constraint from the observed $\omega_f$ assuming the CR is due solely to two identical surface states at the surface and interface, with solid orange curves indicating the error bars. Dashed orange and green curves show the effect of breaking this assumption and allowing 25\% of the Drude weight to be attributed to the confined conduction band states. (b) Possible realizations of a minimal surface state model in \cite{liu10} which are consistent with our observations for different choices of $A$ and $C_2$. The Fermi level $\mu_F$, shown in black, intersects the surface and confined bulk conduction subbands. (c) Dispersion of the surface state quasiparticles determined in this work with 75\% surface state signal, $A$ = 25 meV$\cdot$nm, and $C_2$ = 890 meV$\cdot$nm$^2$.}
\label{fig:band}
\end{center}
\end{figure}

Before we address the implications of possible contributions from the bulk, we start with the simplest assumption, namely that the only contributions to the observed Kerr rotation originate from the two surfaces of the film with approximately the same charge carrier concentration. This is motivated by the magneto-transport showing a quantized Hall effect that results from the 2D Dirac-like topological surface states with densities 4.8$\times$10$^{11}$cm$^{-2}$ and 3.7$\times$10$^{11}$cm$^{-2}$ for the CdTe and vacuum interfaces (respectively) and negligible contribution from the bulk \cite{brune11}. Since with this assumption $\hbar\omega_{f}^j=39.5\pm 0.9$ meV for the two surfaces, combination with $\omega_c/B$ gives $v_F^s=$5.88$\times10^5$ m/s and $k_F^s=0.201$ nm$^{-1}$ (see Fig \ref{fig:band}a). The corresponding carrier concentration per surface is $n_{2D}=k_F^2/(4\pi)=3.2\times 10^{11}$cm$^{-2}$ in good agreement with the high field transport data \cite{brune11} and strongly evidencing a surface-dominated origin of the CR signal. The remaining difference is within the observed variations from one cool-down to another of the same sample, which we attribute to molecular adsorption at the surface of the film at cryogenic temperatures. Comparing these $k_F$ values with the quantum well bandstructure in \cite{brune11}, the level of the chemical potential should be positioned above the conduction band bottom with 25\% of the Drude weight arising from the bulk states of the film, with the remainder due to the surface bands. Attributing only 75\% of the observed Drude weight to the surface states would result in: $v_F^s=$5.09$\times10^5$ m/s and $k_F^s=0.174$ nm$^{-1}$, as indicated by the green dotted lines in Fig. \ref{fig:band}a. Angle-resolved photoelectron spectroscopy shows a linear dispersion down to 1 eV below the Fermi energy with a velocity 4.3$\times10^5$ m/s \cite{brune11}. The somewhat higher value obtained from the optical data is expected, since this probes the velocity at the Fermi energy which is closer to the light conduction bands hybridizing with the 2D surface states.

%
%


A more detailed perspective of the surface-dominated cyclotron resonance can be attained through comparison of these results to a surface state model dispersion relation $\epsilon_k=E_{DP}+Ak+C_2k^2$ \cite{liu10}. This form accommodates the topologically protected Dirac point (DP) through the $A$ term, but allows for significant deviations from the ideal conical dispersion through the $C_2$ term. By varying these parameters, one can interpolate continuously between a pure Dirac cone ($C_2 \rightarrow 0$) and a parabola ($A\rightarrow 0$) emanating from the Dirac point at zone center and energy $E_{DP}$ lying inside the bulk gap. All of the curves of this type with the same values of $k_F$ and $v_F$ fit equally well to our data, as shown in Figure \ref{fig:band}b. However, in order to place the DP inside the bulk gap, estimated to be $\sim26$ meV at zone center \cite{brune11}, the $A$ parameter must be very small, $A<$ 50 meV$\cdot$nm, to be consistent with our data. The resultant parameter $C_2$ is therefore quite large in comparison to other topological materials \cite{hsieh09,hsieh08,chen09,liu10}, indicating a relatively rapid departure from the ideal conical dispersion near the DP in strained HgTe. The large second-order term leads one to suspect that higher order isotropic terms ($k^3$, $k^4$, ...) may be necessary for detailed analysis of certain experiments.



While we do not measure the Fermi energy directly, the simplified dispersion analysis above puts the chemical potential within 40 meV of the Dirac point. This difference can be overcome through gating in an appropriate experimental design, and the chemical potential can be tuned near the Zeeman split zero Landau level. This would fulfill the conditions required to observe the predicted topological magnetoelectric coupling effect \cite{qhz08,tse10a} as a quantized Kerr rotation. This long-sought effect bears close mathematical analogy to high-energy particle theory, permitting one to use terahertz spectroscopy to study the properties of an `axion' domain wall \cite{wilcek87}, and the methods developed here are highly suited for this pursuit. 

We have studied the low energy electrodynamics of topological surface states of strained HgTe using a novel time-domain magneto-optical spectroscopic technique providing Kerr angle spectra to very low energies ($<$1 meV). The method allowed us to obtain the parameters describing the topological surface states near the Fermi energy, until now not resolved by other experimental techniques, namely free carrier spectral weight, quasi-particle scattering rate, cyclotron frequency, Fermi-velocity and Fermi-momentum. Taken together with the requirement that the Dirac point must lie inside the bulk gap, our results imply that the surface bands of strained HgTe are markedly nonconical. 

We acknowledge valuable discussions with Shoucheng Zhang and Alberto Morpurgo. This work is supported by the SNSF through Grant No. 200020-135085 and the National Center of Competence in Research (NCCR) MaNEP.


\bibliography{topo}

\end{document}


\title{Surface state charge dynamics of a three dimensional topological insulator (Supplemental Materials)}


\author{Jason N. \surname{Hancock}$^{1}$, J. L. M. van Mechelen$^{1}$, Alexey B. Kuzmenko$^{1}$, Dirk van der Marel$^{1}$, C. Br\"une$^{2}$, E. G. Novik$^{2}$, G. V. Astakhov$^{2}$, H. Buhmann$^{2}$, Laurens Molenkamp$^{2}$}
\affiliation{$^{1}$ 
D\'epartement de Physique de la Mati\`ere Condens\'ee, Universit\'e de Gen\`eve, quai Ernest-Ansermet 24, CH 1211 Gen\`eve 4, Switzerland}
\affiliation{$^{2}$ 
Physikalisches Institut der Universit\"at W\"urzburg - 97074 W\"urzburg, Germany}

\date{\today}

\maketitle



\subsection{Materials and methods}
High mobility, 70 nm HgTe film sample was grown on a thick CdTe substrate using molecular beam epitaxy, and characterized by X-ray, ARPES, and d. c. transport measurements. The finite area of the film, 4$\times$5 mm, limits the lowest measurable frequency in our measurements. Time-domain terahertz traces were collected with a commercial system employing a mode-locked laser and photoconducting antennae. 

\subsection{Magneto-optical THz measurements}
Equation (1) of the main text is an example of Malus' law, and is straightforward to derive. 

For a given angular momentum state of the photon (+ or -), the transmission coefficient is
\begin{equation}\label{ }
t_{1,\pm}=t_{vs}e^{i\phi_s}t_{sfv,\pm}
\end{equation}
for the first pulse and
\begin{equation}\label{ }
t_{2,\pm}=t_{vs}e^{3i\phi_s}r_{sfv,\pm}r_{sv}t_{sfv,\pm}
\end{equation}
for the second pulse, where 
\begin{equation}
\label{ }
t_{vs}=\frac{2}{n_s+1}
\end{equation}
\begin{equation}
\label{ }
r_{sv}=\frac{n_s-1}{n_s+1}
\end{equation}
are the usual Fresnel coefficients for transmission and reflection at the substrate/vacuum interface, $\phi_s$ is the complex phase shift acquired through traversal across the substrate and $s$, $f$,  and $v$ stand for substrate, film, and vacuum, respectively. 

\begin{figure}
\begin{center}
\includegraphics[width=3.6in]{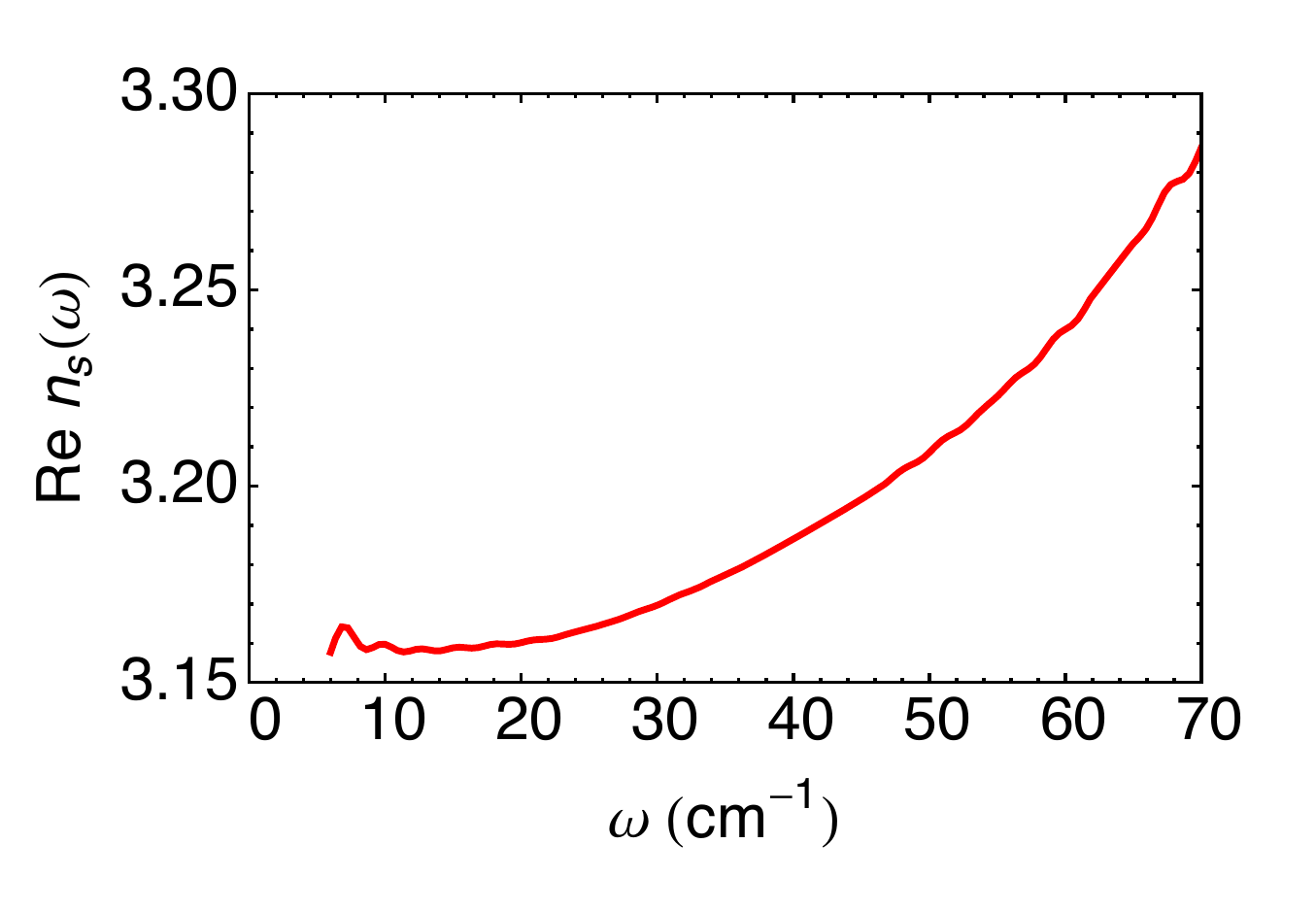}
\caption{Index of refraction of a CdTe substrate measured using terahertz spectroscopy and used to fit the normal incidence Kerr angle $\theta_K$ to the cyclotron resonance formulae (Eqn 2, main text).}
\label{fig:s1}
\end{center}
\end{figure}

\begin{figure}
\begin{center}
\includegraphics[width=3.6in]{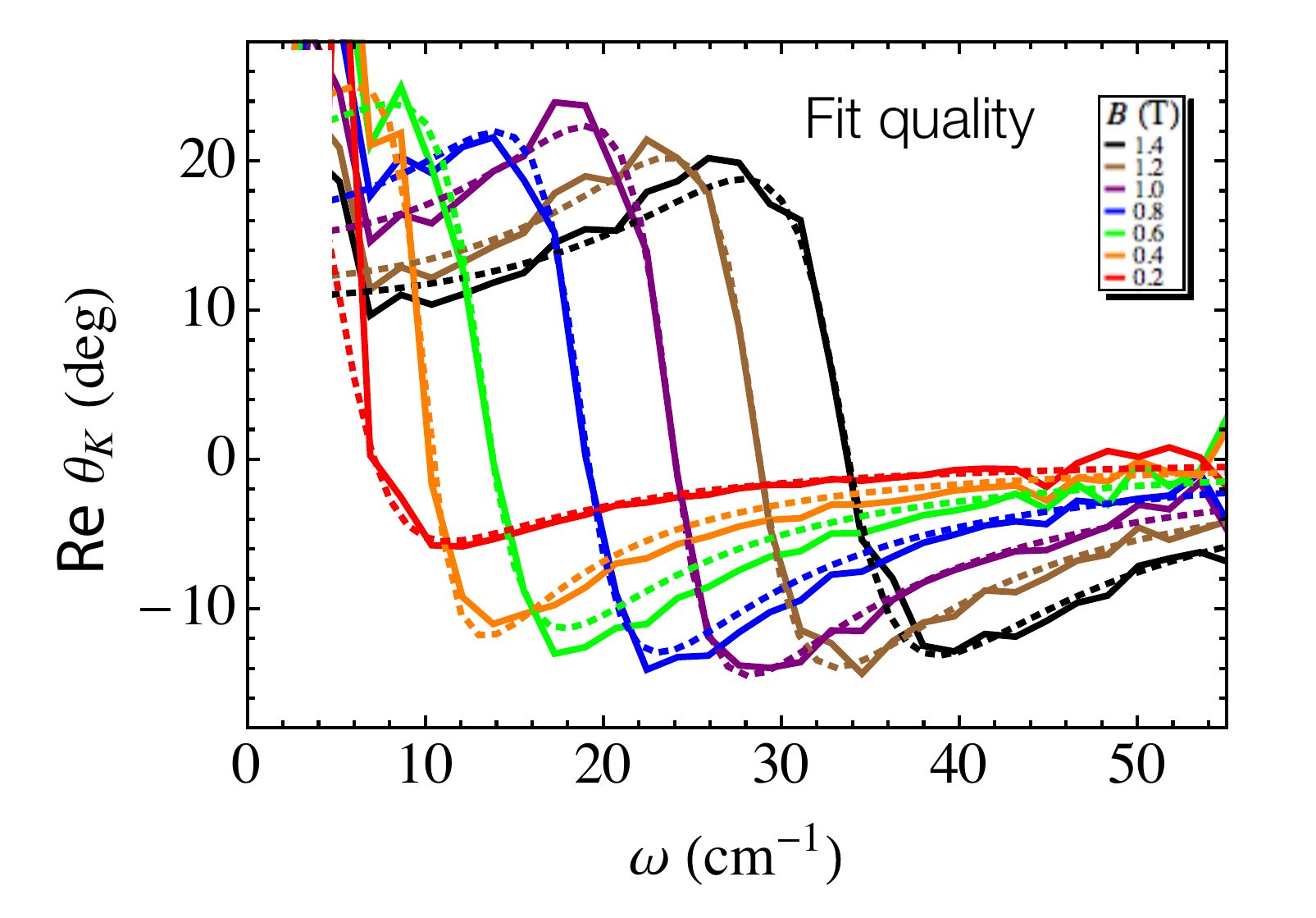}
\caption{Fit quality of the CR formula to the measured Re$\theta_K$ for positive fields.}
\label{fig:s2}
\end{center}
\end{figure}

The coefficients which include the thin film are 
\begin{equation}
\label{ }
t_{sfv,\pm}=\frac{2n_s}{n_s+1+Z_0\sigma_{\pm}}
\end{equation}
\begin{equation}
\label{}
r_{sfv,\pm}=\frac{n_s-1-Z_0\sigma_{\pm}}{n_s+1+Z_0\sigma_{\pm}}
\end{equation}
where $Z_0=\sqrt{\mu_0/\epsilon_0}\simeq376.63$... $\Omega$ is the impedance of free space. The Faraday angle of the $i$th pulse $\theta_{F,i}$ is
\begin{equation}
\label{eq:last}
\tan{\theta_{F,i}}=-i\frac{t_{i,+}-t_{i,-}}{t_{i,+}+t_{i,-}}
\end{equation}
Using a trigonometric identity, 
\begin{equation}
\label{ }
\tan{\theta_K}=\tan{(\theta_{F,2}-\theta_{F,1})}=-\frac{iZ_0n_s(\sigma_+-\sigma_-)}{n_s^2-(1+Z_0\sigma_+)(1+Z_0\sigma_-)}
\end{equation}
This relation connects the model conductivity to the measured normal-incidence Kerr angle. The frequency-dependent index of refraction of the substrate was determined from separate measurements and is shown in Figure \ref{fig:s1}. The quality of fit used to deduce $\omega_c$, $\omega_f$, and  $\gamma$ is shown for positive fields in Figure \ref{fig:s2}.

\subsection{Drude weight and cyclotron frequency, general considerations}
For isotropic planar dispersion, the Drude weight of a band $i$ is
\begin{equation}
\label{}
\omega_f^i=\frac{1}{2\pi}\oint_{FS}\frac{v_F^i(k)^2}{|\vec{v}_F^i(k)|}dk=\frac{1}{2\pi}\frac{v_F^i}{2}(2\pi k_F^i)=\frac{1}{2}v_F^ik_F^i
\end{equation}
The frequency $\omega$ of electrons executing closed orbits of $k$-space area $A$ satisfies \cite{ashcroft}
\begin{equation}
\label{ }
\frac{1}{B} = \frac{2 \pi e}{\hbar^2 \omega} \frac{\partial \epsilon}{\partial A} n
\end{equation}
where the cyclotron frequency $\omega_c$ corresponds to the first harmonic ($n=1$).
 
In our case of an isotropic 2D Fermi surface,
\begin{equation}
\label{ }
 A=\pi k_F^2
\end{equation}
We use the fact that $d \epsilon/dk = \hbar v$ is the group velocity, so that
\begin{equation}
\label{ }
\frac{\partial A}{\partial \epsilon} = 2\pi \frac{k_F}{\hbar v_F}
\end{equation}
from which 
\begin{equation}
\label{ }
\hbar\omega_c = e B \frac{v_F}{k_F}.
\end{equation}

\subsection{Surface state model}

The surface state model is given in reference \cite{liu10} as
\begin{equation}
\label{ }
\epsilon_s(k)=E_{DP}+Ak+C_2k^2.
\end{equation}
The Fermi wavevector is
\begin{equation}
\label{ }
k_{F}^s=\frac{-A+\sqrt{A^2+4C_2(\mu_F-E_{DP})}}{2C_2}.
\end{equation}
and the Fermi velocity is
\begin{equation}
\label{ }
\hbar v_{F}^s=A+2C_2k_{F}^s.
\end{equation}


\bibliography{topo}